\documentclass[pra,twocolumn,showpacs]{revtex4}
\begin{document}
\title{Comment on ``Some implications of the quantum nature of
laser fields for quantum computations''}
\date{January 16, 2003}
\author{Wayne M. Itano}
\affiliation{Time and Frequency Division, National Institute of
Standards and Technology, Boulder, Colorado 80305}
\begin{abstract}
A recent discussion of quantum limitations to the fidelity with which
superpositions of internal atomic energy levels can be generated by an applied,
quantized, laser pulse is shown to be based on unrealistic physical assumptions.
This discussion assumed the validity of Jaynes-Cummings dynamics for an atom
interacting with a laser field in free space, that is, when the atom is not
surrounded by a resonant cavity.  If the laser field is a multimode quantum
coherent state, and the Rabi frequency is much greater than the spontaneous decay
rate, then the total atomic decoherence rate is on the order of the spontaneous
decay rate. With the use of a unitary transformation of the field states due to
Mollow, it can be shown that the atomic decoherence rate is the same as if the
laser field were treated classically, without any additional contribution due to
the quantum nature of the laser field.
\end{abstract}
 \maketitle
 The quantum dynamics of a two-level atom in free space
interacting with a resonant, coherent, quantized electromagnetic field is
important from the standpoint of pure physics and potentially for practical
applications. For example, some proposed implementations of quantum computation
depend on the ability to accurately generate arbitrary superpositions of two
atomic states by means of applied, resonant fields. If the external field is
considered to be classical, then the atomic dynamics (Rabi flopping), including
decoherence due to technical imperfections in the classical driving field, can
easily be calculated [see, e.g., Sec. 4 of Ref. \cite{wineland98}]. In addition,
decoherence due to radiative decay of the atomic states has been considered [see,
e.g., Secs. 4.2.1, 4.4.6.4 of Ref. \cite{wineland98}]. Conceivably, the quantum
nature of the driving field might lead to additional decoherence.

A recent attempt \cite{geabanacloche02} to extend the calculations of Rabi
flopping in free space to the case of a quantized driving field used an
inaccurate model, which is equivalent to a ``reversed micromaser.'' That is,
instead of an atom passing through a resonant cavity, an atom is intercepted by
an electromagnetic field, confined to a region of space travelling at the speed
of light. The context of this calculation was the necessity, in quantum
computation, for high accuracy of quantum state control.  Others have applied a
more or less equivalent model to problems in quantum information processing
\cite{vanenk02}. In the ``reversed micromaser'' model, Fock states
$\left|n\right\rangle$ apparently represent quantized field excitations confined
to an imaginary box moving at the speed of light. While the atom is inside the
field region, the atom-field state is presumed to follow Jaynes-Cummings dynamics
\cite{jaynes63}. In this model, a coherent laser pulse is represented by a
superposition of moving Fock states
$\left|\alpha\right\rangle=e^{-|\alpha|^2}\sum_{n=0}^{\infty}
\left(\alpha^n/\sqrt{n!}\right)\left|n\right\rangle$. Jaynes-Cummings dynamics
then lead to entanglement of the atom and field and to effective decoherence of
the atomic dynamics when a trace is performed over the field degrees of freedom.

This picture is unrealistic and inaccurate for an atom in free space, since there
the field is not confined by a cavity. The physical problem with the
Jaynes-Cummings model in free space is that it assumes that there is only one
mode of the field.  All radiation emitted by the atom must go into that mode, and
all radiation absorbed by the atom must come out of that mode. Thus, emitted
radiation stays around and can be reabsorbed, and the absorption of radiation by
the atom decreases the intensity of the applied field. The combination of these
two effects leads to the complicated Jaynes-Cummings atomic dynamics, including
the well-known collapses and revivals. The former effect (reabsorption of emitted
radiation) does not occur in free space, because the emitted photon leaves the
atom and does not interact with it again.  The latter effect (a decrease in the
applied field upon absorption of radiation by the atom) also does not occur in
free space. It would correspond to a change in the laser pulse amplitude
\emph{upstream} from the atom. A change in the amplitude \emph{downstream} does
of course occur, due to interference with the coherent forward-scattered field.
Radiation is emitted by the atom in a dipole (or other multipole) pattern into
all modes of the field and also as coherent forward scattering. Because the
electromagnetic field has all modes available to it, not just a single one, the
atomic dynamics will differ from those predicted by the Jaynes-Cummings model.

The Jaynes-Cummings model makes an odd prediction, which might be called the
``beam area paradox.''  The Jaynes-Cummings (or ``reversed micromaser'') model
predicts that the decoherence of the atomic system scales inversely with the mean
number of photons $\langle n\rangle$ in the laser pulse. If one keeps the
intensity at the site of the atom constant, but increases $\langle n\rangle$ by
increasing the cross-sectional area of the beam, the decoherence is predicted to
decrease. This has the appearance of being a \emph{nonlocal} effect of the
presence or absence of the field at arbitrarily large distances from the atom.
This result is more explicit in the work of van Enk and Kimble \cite{vanenk02},
where the beam area $A$ appears explicitly in, for example, Eq. (31), and where
they state, ``Decreasing the focal area $A$ will increase the amount of
entanglement.''

If the applied laser field is treated classically, but a phenomenological decay
rate $\gamma$ for the upper level is included, one finds that the atomic
decoherence rate is on the order of $\gamma$ if the field is strong. ``Strong''
here means that the time required for the atom to undergo an induced transition
(Rabi-flop) is much less than the spontaneous lifetime of the upper state. The
perhaps surprising fact is that no additional decoherence of the atomic system
appears when the electromagnetic field is treated quantum mechanically, with the
driving field being a quantized coherent state and the quantized vacuum field
being present to induce spontaneous decay. This can be seen by making use of
Mollow's unitary transformation [Eq. (2.8) of Ref. \cite{mollow75}]. It turns out
that, in its effect on an atom, a quantum coherent field is equivalent to a
classical ($c$-number) field plus the quantum field, initially in the vacuum
state. The proof of this result is given in detail in a textbook \cite{cohen}.
Since this result holds for a multimode coherent field, and not simply for an
infinite plane wave, it is capable of describing a finite travelling laser pulse.
This situation is similar to one that occurs in the calculation of the spectrum
of resonance fluorescence. Mollow's original 1969 calculation of the spectrum
simply \emph{assumed} that the incident field was classical (i.e., $c$-number)
\cite{mollow69}.  In his 1975 calculation \cite{mollow75}, he showed that this
assumption was unnecessary and that the same spectrum is obtained if the incident
field is treated as a quantized coherent state. It should be noted that this is
an exact result, not one that is valid only in the large quantum number
(classical) limit. The initially empty modes of the quantized field are
eventually populated, but only at the timescale of the spontaneous decay. In
addition, there is a coherent, forward-scattered, $c$-number field [see p. 1920
of Ref. \cite{mollow75}] that, being $c$-number, does not lead to entanglement or
decoherence of the atom. The fact that the \emph{total} decoherence rate is of
the order of $\gamma$ follows immediately from the fact that the probability that
the field remains in the vacuum state is $e^{-\gamma t/2}$, where $t$ is the time
after the interaction has has started [Eq. (4.30) of Ref. \cite{mollow75}].  One
can transform back to the ordinary frame, by using the inverse transformation,
but this is not necessary for calculation of the \emph{atomic} decoherence rate,
since the unitary transformation involves only field operators and leaves the
atomic state invariant. There is no ``beam area paradox'' in this treatment,
since the interaction Hamiltonian depends (in the electric dipole approximation),
only on the electric field at the position of the atom [e.g., Eq. (3.2a) of Ref.
\cite{mollow75}]. Even if we go beyond the electric dipole approximation, the
interaction still depends only on local properties, such as derivatives, of the
field.

The main conclusion is that the decoherence of the atomic state upon application
of a quantized, coherent field can be made as small as desired by making the
interaction time sufficiently short compared to the spontaneous decay time.  Of
course, the intensity of the applied field must be high enough so that the
desired operation, such as a $\pi$ transition, can be carried out in that time.
The total decoherence rate is of the order of $\gamma$. There is no additional
decoherence due to the quantum nature of the applied field, as long as it is in a
coherent state. A similar conclusion holds for a Raman transition in a multilevel
atom. That is, if the applied fields are coherent, then decoherence is the same
as if the applied fields were classical and can be attributed to spontaneous
emission [see, e.g., Sec. 4.4.6.4 of Ref. \cite{wineland98} for a discussion of
decoherence for Raman transitions driven with classical fields].

I thank  I. H. Deutsch for bringing to my attention the fact that
he and A. Silberfarb have independently reached similar
conclusions \cite{silberfarb02}. Some confusion might arise from
the fact that, in the second paragraph of Sec.~II of
Ref.~\cite{silberfarb02}, Silberfarb and Deutsch state, regarding
Refs.~\cite{geabanacloche02,vanenk02}, that ``their conclusions
are correct,'' but that ``one must take great care to understand
the regimes under which this formalism is applicable.''  In the
rest of the paragraph, it is made clear that the formalism is
\emph{not} applicable to precisely the case under question, that
is, to an atom in free space.  They explicitly criticize the use
of the Jaynes-Cummings Hamiltonian [their Eq.~(1)], which
``falsely predicts the possibility of a single photon $2\pi$ pulse
in free space, whereby the photon is perfectly absorbed and
re\"emitted into the original mode.''  They further criticize the
solutions for violating causality.  In the following paragraph
they trace the problems with causality to a ``faulty quantization
procedure.''

In a recent preprint [Sec. III of Ref. \cite{geabanacloche02b}], Gea-Banacloche
modifies the arguments of Ref. \cite{geabanacloche02} and claims that it is
really the number of photons $n^\prime$ within a certain volume that is important
for the decoherence, not the total number of photons $n$ in the laser pulse. That
volume is given by the product of an effective cross section $\sigma_{\rm eff}$
and the length of the laser pulse. The effective cross-section is $\sigma_{\rm
eff}=3\pi/2k^2$, where $k$ is the wavenumber of the incident light.  Even if this
result has the right order of magnitude, as it appears to, the definition of
$n^\prime$ seems to be arbitrary and seems to have been chosen to give the
desired result.

This work was supported by the U. S. National Security Agency
(NSA) and Advanced Research and Development Activity (ARDA) under
Contract No. MOD-7171.00 and the U. S. Office of Naval Research
(ONR).  This work is a contribution of NIST, an agency of the U.
S. government, and is not subject to U. S. copyright.

\end{document}